\documentclass[modern]{aastex61}
\usepackage{xspace}
\usepackage[utf8]{inputenc}
\usepackage[T1]{fontenc}
\usepackage{ulem}

\graphicspath{{figures/}}

\submitjournal{ApJ}

\shorttitle{Astropy Project II}
\shortauthors{The Astropy Collaboration}

\newcommand{\package}[1]{\texttt{#1}\xspace}
\newcommand{\github}{\package{GitHub}}
\newcommand{\python}{\package{Python}}
\newcommand{\astropy}{Astropy\xspace}
\newcommand{\astropypkg}{\package{astropy}}

\newcommand{\sectionname}{Section\xspace}
\renewcommand{\figurename}{Figure\xspace}

\renewcommand{\tablename}{Table\xspace}

\usepackage[colorinlistoftodos]{todonotes}

\begin{document}

\draft{\today}

\title{The Astropy Project: Building an inclusive, open-science project and status of the v2.0 core package}

\correspondingauthor{Astropy Coordination Committee}
\email{coordinators@astropy.org}

\author{The Astropy Collaboration}
\noaffiliation
{\let\thefootnote\relax\footnote{{The author list has three parts: the authors that made significant contributions to the writing of the paper in order of contribution, the four members of the \astropy Project coordination committee, and contributors to the \astropy Project in alphabetical order. The position in the author list does not correspond to contributions to the \astropy Project as a whole. A more complete list of contributors to the core package can be found in the \href{https://github.com/astropy/astropy/graphs/contributors}{package repository}, and at the \href{http://www.astropy.org/team.html}{\astropy team webpage}.}}}

\newcommand{\afprinceton}{Department of Astrophysical Sciences, Princeton University, Princeton, NJ 08544, USA}
\newcommand{\afstsci}{Space Telescope Science Institute, 3700 San Martin Dr., Baltimore, MD 21218, USA}
\newcommand{\afsaao}{South African Astronomical Observatory, PO Box 9, Observatory 7935, Cape Town, South Africa}
\newcommand{\afminnstate}{Department of Physics and Astronomy, Minnesota State University Moorhead, 1104 7th Ave S, Moorhead, MN 56563}
\newcommand{\afgoddard}{NASA Goddard Space Flight Center, 8800 Greenbelt Road, Greenbelt, MD 20771, USA}
\newcommand{\afcfa}{Harvard-Smithsonian Center for Astrophysics, 60 Garden St., Cambridge, MA, 02138, USA}
\newcommand{\afwesternontario}{Department of Physics \& Astronomy, University of Western Ontario, 1151 Richmond St, London ON N5X4H1 Canada}
\newcommand{\afnasaames}{NASA Ames Research Center, Moffett Field, CA 94043, USA}
\newcommand{\afjhu}{Department of Physics and Astronomy, Johns Hopkins University, Baltimore, MD 21218, USA}
\newcommand{\afmpik}{Max-Planck-Institut f\"ur Kernphysik, PO Box 103980, 69029 Heidelberg, Germany}
\newcommand{\afioa}{Institute of Astronomy, University of Cambridge, Madingley Road, Cambridge, CB3 0HA, UK}
\newcommand{\afpennstate}{Dept of Astronomy and Astrophysics, Pennsylvania State University, University Park, PA 16802}
\newcommand{\afgemini}{Gemini Observatory, 670 N. Aohoku Pl, Hilo, HI 96720, USA}
\newcommand{\aflco}{Las Cumbres Observatory, 6740 Cortona Drive, Suite 102, Goleta, CA 93117-5575, USA}
\newcommand{\afucsb}{Department of Physics, University of California, Santa Barbara, CA 93106-9530, USA}
\newcommand{\afesomunich}{European Southern Observatory, Karl-Schwarzschild-Stra{\ss}e 2, 85748 Garching bei M\"{u}nchen, Germany}
\newcommand{\afuw}{Department of Astronomy, University of Washington, Seattle, WA 98155}
\newcommand{\afberkeleyastro}{Department of Astronomy, UC Berkeley, 501 Campbell Hall \#3411, Berkeley, CA 94720, USA}
\newcommand{\afuct}{Department of Astronomy, University of Cape Town, Private Bag X3, Rondebosch 7701, South Africa}

\author[0000-0003-0872-7098]{A. M. Price-Whelan}
\affiliation{\afprinceton}

\author[0000-0002-3713-6337]{B. M. Sip\H{o}cz}
\noaffiliation

\author[0000-0003-4243-2840]{H. M. G\"{u}nther}
\affiliation{Kavli Institute for Astrophysics and Space Research, Massachusetts Institute of Technology, 70 Vassar St., Cambridge, MA 02139, USA}

\author[0000-0003-0079-4114]{P. L. Lim}
\affiliation{\afstsci}

\author[0000-0002-8969-5229]{S. M. Crawford}
\affiliation{\afsaao}

\author[0000-0002-3657-4191]{S. Conseil}
\affiliation{Univ Lyon, Univ Lyon1, Ens de Lyon, CNRS, Centre de Recherche Astrophysique de Lyon UMR5574, F-69230, Saint-Genis-Laval, France}

\author[0000-0003-4401-0430]{D. L. Shupe}
\affiliation{Caltech/IPAC, 1200 E. California Blvd, Pasadena, CA 91125}

\author[0000-0001-7988-8919]{M. W. Craig}
\affiliation{\afminnstate}

\author[0000-0002-5686-9632]{N. Dencheva}
\affiliation{\afstsci}

\author[0000-0001-6431-9633]{A. Ginsburg}
\affiliation{National Radio Astronomy Observatory, 1003 Lopezville Rd, Socorro, NM 87801}

\author[0000-0002-9623-3401]{J. T. VanderPlas}
\affiliation{eScience Institute, University of Washington, 3910 15th Ave NE, Seattle, WA 98195, USA}

\author[0000-0002-7908-9284]{L. D. Bradley}
\affiliation{\afstsci}

\author[0000-0003-0784-6909]{D. P\'{e}rez-Su\'{a}rez}
\affiliation{University College London/Research IT Services, Gower St, Bloomsbury, London WC1E 6BT, United Kingdom}

\author[0000-0002-0455-9384]{M. de Val-Borro}
\affiliation{Astrochemistry Laboratory, \afgoddard}

\collaboration{(primary paper contributors)}

\author{T. L. Aldcroft}
\affiliation{\afcfa}

\author[0000-0002-1821-0650]{K. L. Cruz}
\affiliation{Department of Physics and Astronomy, Hunter College, City University of New York, 695 Park Avenue, New York, NY 10065}
\affiliation{Physics, Graduate Center of the City University of New York, New York, NY, USA}
\affiliation{Department of Astrophysics, American Museum of Natural History, New York, NY, USA}

\author[0000-0002-8642-1329]{T. P. Robitaille}
\affiliation{Aperio Software Ltd., Headingley Enterprise and Arts Centre, Bennett Road, Leeds, LS6 3HN, United Kingdom}

\author[0000-0002-9599-310X]{E. J. Tollerud}
\affiliation{\afstsci}

\collaboration{(Astropy coordination committee)}

\author{C. Ardelean}
\affiliation{\afwesternontario}

\author[0000-0002-8222-3595]{T. Babej}
\affiliation{Department of Theoretical Physics \& Astrophysics, Masaryk University, Kotlarska 2, 61137 Brno, Czech Republic}

\author[0000-0002-4576-9337]{M. Bachetti}
\affiliation{INAF-Osservatorio Astronomico di Cagliari, via della Scienza 5, I-09047, Selargius, Italy}

\author{A. V. Bakanov}
\noaffiliation

\author[0000-0001-7821-7195]{S. P. Bamford}
\affiliation{School of Physics \& Astronomy, University of Nottingham, University Park, Nottingham NG7 2RD, UK}

\author[0000-0002-3306-3484]{G. Barentsen}
\affiliation{\afnasaames}

\author[0000-0003-2767-0090]{P. Barmby}
\affiliation{\afwesternontario}

\author[0000-0002-9374-2729]{A. Baumbach}
\affiliation{Heidelberg University, Kirchhoff Institut for Physics, Im Neuenheimer Feld 227, 69116 Heidelberg, Germany}

\author{K. L. Berry}
\noaffiliation

\author{F. Biscani}
\affiliation{Max-Planck-Institut f\"ur Astronomie, K\"onigstuhl 17, 69117 Heidelberg, Germany}

% ...

\collaboration{(Astropy contributors)}

\begin{abstract}
The \astropy project supports and fosters the development of open-source and openly-developed
\python packages that provide commonly-needed functionality to the astronomical
community.
A key element of the \astropy project is the core package \astropypkg, which serves as the
foundation for more specialized projects and packages.
In this article, we provide an overview of the organization of the \astropy
project and summarize key features in the core package as of the recent major
release, version 2.0.
We then describe the project infrastructure designed to facilitate and support
development for a broader ecosystem of inter-operable packages.
We conclude with a future outlook of planned new features and directions for the
broader \astropy project.
\end{abstract}


\section{Introduction} \label{sec:intro}
All modern astronomical research makes use of software in some way.
Astronomy as a field has thus long supported the development of software tools
for astronomical tasks, such as scripts that enable individual scientific
research, software packages for small collaborations, and data reduction
pipelines for survey operations.
Some software packages are, or were, supported by large institutions and are
intended for a wide range of users.
These packages therefore typically provide some level of documentation and user
support or training.
Other packages are developed by individual researchers or research groups and
are then typically used by smaller groups for more domain-specific purposes.
For both packages meant for wider distribution and for scripts specific to
particular research projects, a library that addresses common astronomical tasks
simplifies the software development process.
The users of such a library then also benefit from a community and ecosystem
built around a shared foundation.
The \astropy project has grown to become this community for \python astronomy
software, and the \astropypkg core package is a feature-rich \python library.

The development of the \astropypkg core package began as a largely
community-driven effort to standardize core functionality for astronomical
software in \python.
In this way, its genesis differs from, but builds upon, many substantial and
former astronomical software development efforts that were commissioned or
initiated through large institutional support, such as IRAF \citep[developed
at NOAO;][]{IRAF}, MIDAS \citep[developed at ESO;][]{MIDAS}, or Starlink
\citep[originally developed by a consortium of UK institutions and now
maintained by the East Asian Observatory;][]{starlink1982,starlink2013}.
More recently, community-driven efforts have seen significant success in the astronomical sciences (e.g., \citealt{yt}).

\python\footnote{\url{https://www.python.org/}} is an increasingly popular, general-purpose
programming language that is available under a permissive open source software license and is free of
charge for all major operating systems. The programming language has become especially popular
in the quantitative sciences, where researchers must simultaneously produce research, perform
data analysis, and develop software. A large part of this success owes itself to the vibrant
community of developers and a continuously-growing ecosystem of tools, web services, and stable
well-developed packages that enable easier collaboration on software development, easier
writing and sharing of software documentation, and continuous testing and validation of
software. While dedicated libraries provide support for array representation and
arithmetic \citep[\package{numpy};][]{numpy}, a wide variety of functions for scientific
computing \citep[\package{scipy};][]{scipy}, and publication-quality plotting
\citep[\package{matplotlib};][]{matplotlib}, tens of thousands of other high-quality and easy-to-use
packages are available, which can help with tasks that are not specific to astronomy but
might be performed in the course of astronomical research, e.g., interfacing with
databases, or statistical inferences.
More recently, the development and mainstream adoption of package managers such
as Anaconda\footnote{\url{https://anaconda.org/}} has significantly streamlined
the installation process for many libraries, lowering the barriers to entry.

The \astropy project aims to provide an open-source and open-development core
package (\astropypkg) and an ecosystem of \emph{affiliated packages} that
support astronomical functionality in the \python programming language.
The \astropypkg core package is now a feature-rich library of sufficiently
general tools and classes that supports the development of more specialized
code. An example of such functionality is reading and writing FITS files: It would be
time consuming and impractical for multiple groups to implement the FITS
standard \citep{FITS} and maintain software for such a general-purpose need.
Another example of such a common task is in dealing with representations of and
transformations between astronomical coordinate systems.

The \astropy project aims to develop and provide high-quality code and
documentation according to the best practices in software development.
The project makes use of different tools and web services to reach those
goals without central institutional oversight.
The first public release of the \astropypkg package is described in
\cite{astropy}. Since then, the \astropypkg package has been
used in hundreds of projects and the scope of the package has grown
considerably. At the same time, the scientific community
contributing to the project has grown tremendously and an ecosystem
of packages supporting or affiliated with the \astropypkg core has
developed.
In this paper, we describe the current status of the \astropy community and the
\astropypkg core package and discuss goals for future development.

We start by describing the way the \astropy project functions and is organized
in \sectionname~\ref{sec:org}. We then describe the main software efforts
developed by the \astropy project itself: a core package called \astropypkg
(\sectionname~\ref{sec:core}) and several separate packages that help maintain
the infrastructure for testing and documentation
(\sectionname~\ref{sec:infrastructure}). We end with a short vision for
the future of \astropy and astronomical software in general in
\sectionname~\ref{sec:future}. The full paper, including the code
to produce the figures, is available in a \github
repository\footnote{\url{https://github.com/astropy/astropy-v2.0-paper}}

This article is not intended as an introduction to \astropypkg, nor does it
replace the \astropypkg documentation. Instead, it describes the way the
\astropy community is organized and the current state of the \astropypkg
package.

\section{Organization and infrastructure}
\label{sec:org}

\subsection{Coordination of Astropy}
\label{sect:coordcom}
From its inception, \astropy has required coordination to ensure the project
as a whole and its coding efforts are consistent and reasonably efficient.
While many \python projects adopt a ``Benevolent Dictator For Life'' (BDFL)
model, \astropy has instead opted for a \emph{coordination committee}.  This
is in part due to the nature of the project as a large-scale collaboration
between many contributors with many interests, and in part due to simply the
amount of work that needs to get done.  For the latter reason, the
project has expanded the committee from three to four members starting in
2016.

For resolving disagreements about the \astropypkg core package or other \astropy-managed code, the coordination committee primarily acts to work toward consensus, or when consensus is difficult to achieve, generally acts as a ``tie-breaker.''
The committee also oversees affiliated package applications to ensure that they
are in keeping with \astropy's vision and
philosophy,\footnote{\url{http://docs.astropy.org/en/stable/development/vision.html}}
as well as the associated procedures.
Additionally, the committee oversees the assignment of roles (primarily driven by already-existing contributions), and increasingly has acted as the ``face'' of the Project, providing contact with organizations like NumFOCUS (the body that holds any potential funding in trust for \astropy) and the American Astronomical Society (AAS).

\subsection{Astropy development model}

Code is contributed to the \astropypkg core package or modified through ``pull
requests'' (via \github\footnote{\url{https://github.com/astropy/astropy/}})
that often contain several \texttt{git} commits.
Pull requests may fix bugs, implement new features, or improve or modify the
infrastructure that supports the development and maintenance of the package.
Individual pull requests are generally limited to a single conceptual addition
or modification to make code review tractable.
Pull requests that modify or add code to a specific subpackage must be reviewed
and approved by one of the subpackage maintainers before they are merged into
the core codebase.
Bugs and feature requests are reported via the \github issue tracker and labeled
with a set of possible labels that help classify and organize the issues.
The development workflow is detailed in the \astropypkg
documentation.\footnote{\emph{How to make a code contribution},
\url{http://docs.astropy.org/en/latest/development/workflow/development_workflow.html}}

\begin{figure}
\includegraphics[width=\textwidth]{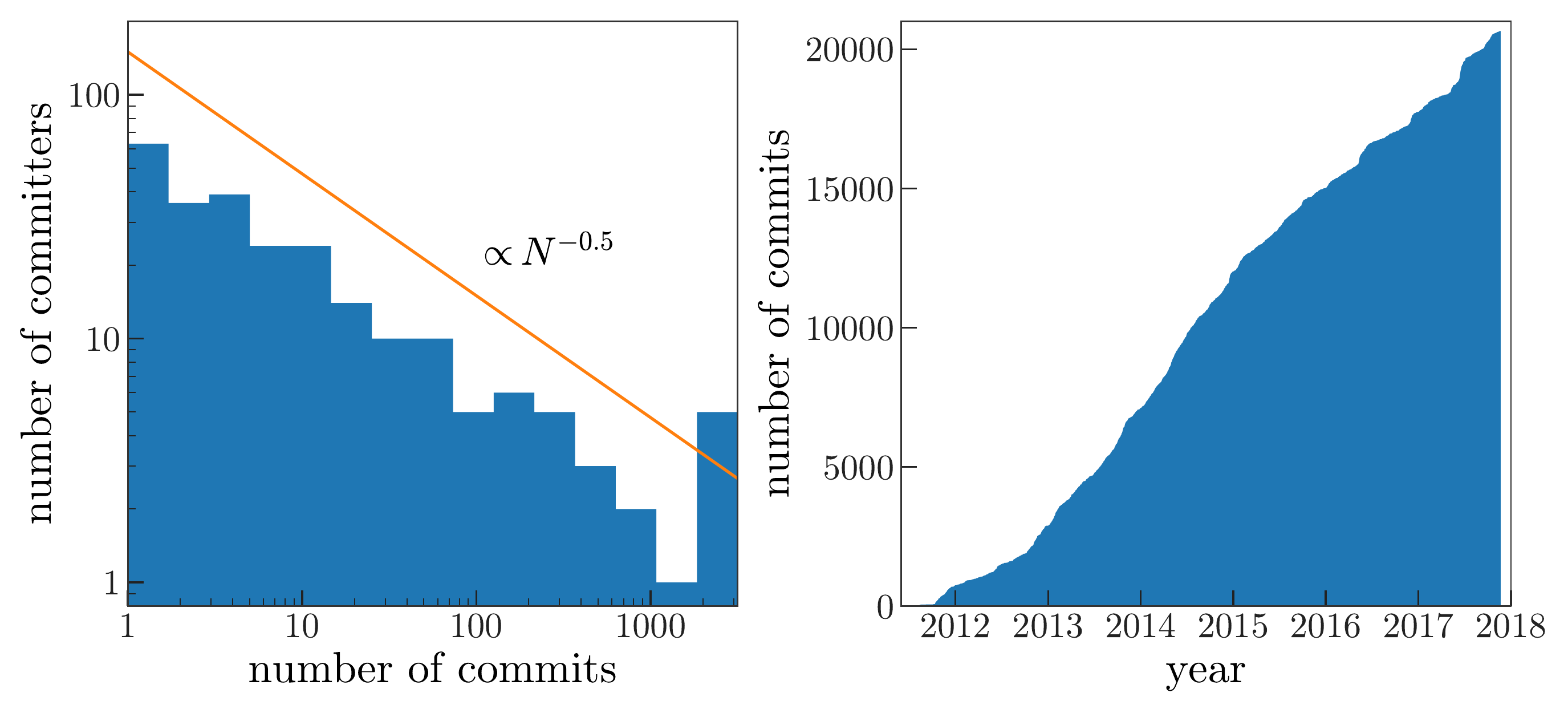}
\caption{%
    \emph{Left panel}: Distribution of number of commits per committer.
    \emph{Right panel}: Cumulative number of commits to the \astropypkg core
    package over time.
    \label{fig:ncommits}
}
\end{figure}

As of version 2.0, \astropypkg contains $212244$ lines of code\footnote{This
line count includes comments, as these are often as important for
maintainability as the code itself.  Without comments there are $142197$ lines
of code.} contributed by $232$ unique contributors over $19270$ \texttt{git}
commits.
\figurename~\ref{fig:ncommits}, left, shows the distribution of total number of
commits per contributor as of November 2017.
The relative flatness of this distribution (as demonstrated by its log-log slope of
$-0.5$) shows that the \astropypkg core package has been developed by a broad
contributor base.  A leading group of 6 developers have each added over 1000
commits to the repository, and $\sim 20$ more core contributors have contributed
at least 100 commits.
However, the distribution of contribution level (number of commits) continues
from 100 down to a single commit.
In this sense, the development of the core package has been a true community
effort and is not dominated by a single individual.
It is also important to note that the number of commits is only a rough metric
of contribution, as a single commit could be a critical fix in the package or a
fix for a typographical error.
\figurename~\ref{fig:ncommits}, right, shows the number of commits as a
function of time since the genesis of the \astropypkg core package.
The package is still healthy: new commits are and have been contributed at a
steady rate throughout its existence.

\subsection{APEs - Astropy Proposals for Enhancement}

Central to the success of \astropy is an open environment where anybody can
contribute to the project.
This model leads to an ``organic'' growth, where features are
implemented by different people with different programming styles and
interfaces.
Thus, \astropy has a mechanism to more formally propose significant changes to
the core package (e.g., re-writing the coordinates subpackage; \citealt{ape5}),
to plan out major new features (e.g., a new file format; \citealt{ape6}), or
institute new organization-wide policies (e.g., adopting a code of conduct;
\citealt{ape8}).
This mechanism is called ``Astropy Proposal for Enhancement'' (APE) and is
modeled after the ``Python Enhancement Proposals'' (PEP) that guide the
development of the \python programming language.
In an APE, one or more authors describe in detail the proposed changes or
additions, including a rationale for the changes, how these changes will be
implemented, and in the case of code, what the interface will be \citep{ape1}.
The APEs are discussed and refined by the community before much work is invested
into a detailed implementation; anyone is welcome to contribute to these
discussions during the open consideration period. APEs are proposed via pull
requests on a dedicated GitHub repository\footnote{\url{https://github.com/astropy/astropy-APEs}};
anyone can therefore read the proposed APEs and leave in-line comments.
When a community consensus emerges, the APEs are accepted and become
the basis for future work.
In cases where consensus cannot be reached, the
\astropy coordination committee may decide to close the discussion and
make an executive decision based on the community input on the APE in question.

\subsection{Concept of affiliated packages}

A major part of the \astropy project is the concept of ``Affiliated Packages.''
An affiliated package is an astronomy-related
\python package that is not part of the \astropypkg core package, but
has requested to be included as part of the \astropy project's
community. These packages support the goals and vision of \astropy of
improving code re-use, interoperability, and embracing good coding
practices such as testing and thorough documentation.

Affiliated packages contain functionality that is more specialized,
have license incompatibilities, or have external dependencies (e.g., GUI
libraries) that make these packages more suitable to be separate from the
\astropypkg core package.
Affiliated packages may also be used to develop substantial new functionality
that will eventually be incorporated into the \astropypkg core package
(e.g., \texttt{wcsaxes}).
New functionality benefits from having a rapid development and release cycle that is not tied to that of the \astropypkg core (\sectionname~\ref{sect:releasecycle}).

Affiliated packages are listed on the main \astropy website and advertised to
the community through \astropy mailing lists; a list of current affiliated
packages is included in \tablename~\ref{tab:registry}.
Becoming an affiliated package is a good way for new and existing packages to
gain exposure while promoting \astropy's high standard for code and
documentation quality.
This process of listing and promoting affiliated packages is one way in which
the \astropy project tries to increase code re-use in the astronomical
community.

Packages can become affiliated to \astropy by applying for this status on a public mailing list. The coordination committee (\sectionname~\ref{sect:coordcom}) reviews such requests and issues recommendations for the improvement of a package, where applicable.

\subsection{Release cycle and Long Term Support}
\label{sect:releasecycle}

The \astropypkg package has a regular release schedule consisting of new significant
releases every 6 months, with bugfix releases as needed \citep{ape2}.
The major releases contain new features or any significant changes, whereas
the bugfix releases only contain fixes to code or documentation but no new
features.
Some versions are additionally designated as ``Long-term support'' (LTS)
releases, which continue to receive bug fixes for 2 years following the release
with no changes to the API\@.
The LTS versions are ideal for pipelines and other applications in which API
stability is essential.
The latest LTS release (version 2.0) is also the last one that supports \python
2; it will receive bug fixes until the end of 2019 \citep{ape10}.

The version numbering of the \astropypkg core package reflects this release
scheme: the core package version number uses the form \texttt{x.y.z}, where ``x'' is
advanced for LTS releases, ``y'' for non-LTS feature releases, and
``z'' for bugfix releases.

The released versions of the \astropypkg core package are available from several
of the \python distributions for scientific computing (e.g.,
\href{http://anaconda.org}{Anaconda}) and from the \python Package Index
(PyPI).\footnote{See the installation documentation for more information:
\url{http://docs.astropy.org/en/stable/install.html}}
Effort has been made to make \astropypkg available and easily installable across
all platforms; the package is constantly tested on different platforms as part
of a suite of continuous integration tests.

\subsection{Support of Astropy}

The \astropy project, as of the version 2.0 release, does not receive any direct
financial support for the development of \astropypkg.
Development of the software, all supporting materials, and community support are
provided by individuals who work on the \astropy project in their own personal
time, by individuals or groups contributing to \astropy as part of a research
project, or contributions from institutions that allocate people to work on
\astropy.
A list of organizations that have contributed to \astropy in this manner
can be found in the Acknowledgments.

Different funding models have been proposed for support of \astropy
(e.g., \citealt{2016arXiv161003159M}), but a long-term plan
for sustainability has not yet been established.
The \astropy project has the ability to accept financial contributions
from institutions or individuals through the NumFOCUS\footnote{NumFOCUS
is a 501(c)(3) nonprofit that supports and promotes world-class, innovative,
open source scientific computing.} organization. NumFOCUS has, to date, covered the direct costs incurred by the \astropy project.

\section{Astropy Core Package Version 2.0}
\label{sec:core}
The \astropy project aims to provide \python-based packages for all tasks that
are commonly needed in a large subset of the astronomical community.
At the foundation is the \astropypkg core package, which provides general
functionality (e.g., coordinate transformations, reading and writing
astronomical files, and units) or base classes for other
packages to utilize for a common interface (e.g., \texttt{NDData}).
In this section, we highlight new features introduced or substantially improved
since version 0.2 (previously described in \citealt{astropy}).  The \astropypkg
package
provides a full log of changes\footnote{\url{https://github.com/astropy/astropy/blob/stable/CHANGES.rst}}
over the course of the entire project and more details about individual
subpackages are available in the documentation.\footnote{\url{http://docs.astropy.org/en/stable/}}
Beyond what is mentioned below, most subpackages have seen improved performance
since the release of the version 0.2 package.

\subsection{Units}\label{sec:units}

The \texttt{astropy.units} subpackage adds support for representing units and
numbers with associated units --- ``quantities'' --- in code.
Historically, quantities in code have often been represented simply as numbers,
with units implied or noted via comments in the code because of considerations
about speed: having units associated with numbers inherently adds overhead to
numerical operations.
In \texttt{astropy.units}, \texttt{Quantity} objects extend \texttt{numpy}
array objects and have been designed with speed in mind.

As of \astropypkg version 2.0, units and quantities, prevalent in most of its
subpackages, have become a key concept for using the package as a whole.
Units are intimately entwined in the definition of astronomical coordinates;
thus, nearly all functionality in the \texttt{astropy.coordinates} subpackage
(see \sectionname~\ref{sec:coordinates}) depends on them.
For most other subpackages, quantities are at least accepted and often expected
by default.

The motivation and key concepts behind this subpackage were described in detail
in the previous paper \citep{astropy}. Therefore, we primarily highlight new
features and improvements here.

\subsubsection{Interaction with \package{numpy} arrays}

\texttt{Quantity} objects extend
\texttt{numpy.ndarray} objects and therefore
work well with many of the functions in \texttt{numpy} that support
array operations. For example, \texttt{Quantity} objects with angular
units can be directly passed in to the trigonometric functions implemented in
\texttt{numpy}. The units are internally converted to radians, which is what
the \texttt{numpy} trigonometric functions expect, before being passed to
\texttt{numpy}.

\subsubsection{Logarithmic units and magnitudes}
        By default, taking the logarithm of
        a \texttt{Quantity} object with non-dimensionless units intentionally
        fails.
        However, some well-known units are actually logarithmic quantities,
        where the logarithm of the value is taken with respect to some reference
        value.
        Examples include astronomical magnitudes, which are logarithmic fluxes,
        and decibels, which are more generic logarithmic ratios of quantities.
        Logarithmic, relative units are now supported in \texttt{astropy.units}.

\subsubsection{Defining functions that require quantities}
        When writing code or
        functions that expect \texttt{Quantity} objects, we often want to
        enforce that the input units have the correct physical type.
        For example, we may want to require only length-type \texttt{Quantity}
        objects.
        \texttt{astropy.units} provides a tool called \texttt{quantity\_input()}
        that can perform this verification automatically to avoid repetitive
        code.

\subsection{Constants}

The \texttt{astropy.constants} subpackage provides a selection of physical and
astronomical constants as \texttt{Quantity} objects (see
\sectionname~\ref{sec:units}).
A brief description of this package was given in \cite{astropy}.
In version 2.0, the built-in constants have been organized into modules for
specific versions of the constant values.
For example, physical constants have \texttt{codata2014} \citep{codata2014} and
\texttt{codata2010} versions.
Astronomical constants are organized into \texttt{iau2015} and \texttt{iau2012}
modules to indicate their sources (resolutions from the International
Astronomical Union, IAU).
The \texttt{codata2014} and \texttt{iau2015} versions are combined into the
default constant value version: \texttt{astropyconst20}.
For compatibility with \astropypkg version 1.3,  \texttt{astropyconst13}
is available and provides access to the adopted versions of the
constants from earlier versions of \astropypkg.
To use previous versions of the constants as \emph{units} (e.g., solar masses),
the values have to be imported directly; with version
2.0, \texttt{astropy.units} uses the \texttt{astropyconst20} versions.

Astronomers using \texttt{astropy.constants} should take particular note of the
constants provided for Earth, Jupiter, and the Sun.
Following IAU 2015 Resolution B3 \citep{iau2015b3}, nominal values are now given
for mass parameters and radii.
The nominal values will not change even as ``current best estimates'' are
updated.

\subsection{Coordinates}
\label{sec:coordinates}
The \package{astropy.coordinates} subpackage is designed to support representing
and transforming celestial coordinates and, new in version 2.0, velocities.
The framework heavily relies on the \package{astropy.units} subpackage, and most
inputs to objects in this subpackage are expected to be \texttt{Quantity}
objects.
Some of the machinery also relies on the Essential Routines of Fundamental
Astronomy (ERFA) \texttt{C} library for some of the critical underlying
transformation machinery \citep{erfa}, which is based on the Standards Of
Fundamental Astronomy (SOFA) effort \citep{sofa}.

A key concept behind the design of this subpackage is that coordinate
\textit{representations} and \textit{reference systems / frames} are independent
of one another.
For example, a set of coordinates in the International Celestial Reference
System (ICRS) reference frame could be represented as spherical (right
ascension, declination, and distance from solar system barycenter) or Cartesian
coordinates ($x$, $y$, $z$ with the origin at barycenter).
They can therefore change representations independent of being transformed to
other reference frames (e.g., the Galactic coordinate frame).

The classes that handle coordinate representations (the \texttt{Representation}
classes) act like three-dimensional vectors and thus support vector arithmetic.
The classes that represent reference systems and frames (the \texttt{Frame}
classes) internally use \texttt{Representation} objects to store the coordinate
data---that is, the \texttt{Frame} classes accept coordinate data, either as a
specified \texttt{Representation} object, or using short-hand keyword arguments
to specify the components of the coordinates.
These preferred representation and short-hand component names differ between
various astronomical reference systems.
For example, in the ICRS frame, the spherical angles are right ascension
(\texttt{ra}) and declination (\texttt{dec}), whereas in the Galactic frame, the
spherical angles are Galactic longitude (\texttt{l}) and latitude (\texttt{b}).
Each \texttt{Frame} class defines its own component names and preferred \texttt{Representation} class.
The frame-specific component names map to corresponding components on the
underlying \texttt{Representation} object that internally stores the coordinate
data.
For most frames, the preferred representation is spherical, although this is
determined primarily by the common use in the astronomical community.

\begin{figure}
\includegraphics[width=\textwidth]{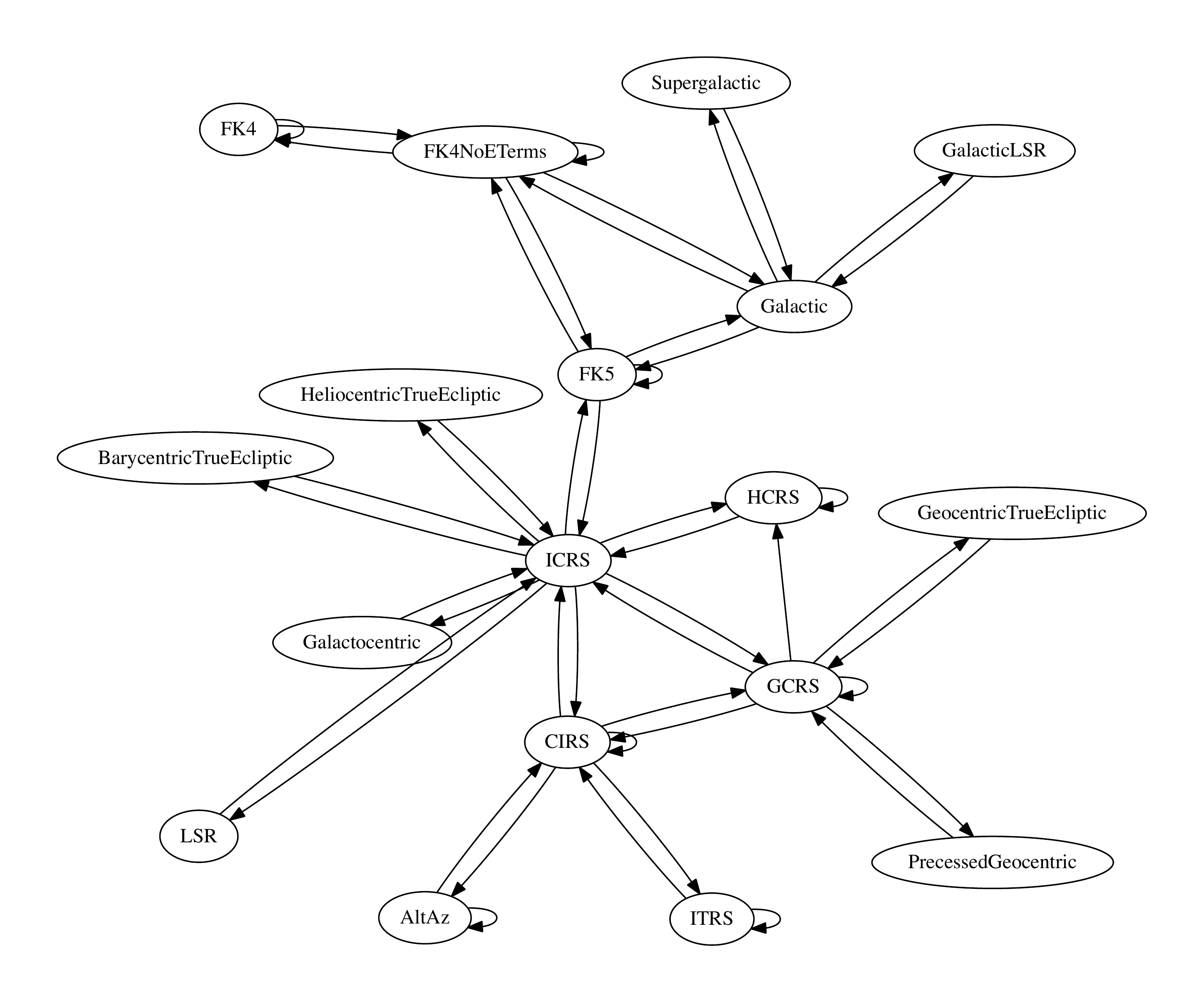}
\caption{%
    The full graph of possible reference frame transformations implemented in
    \texttt{astropy.coordinates}.
    Arrows indicate transformations from one frame to another.
    Arrows that point back to the same frame indicate self-transformations that
    involve a change of reference frame parameters (e.g., equinox).
    \label{fig:frame-transform-graph}
}
\end{figure}

Many of the \texttt{Frame} classes also have attributes specific to the
corresponding reference system that allow the user to specify the frame.
For example, the Fifth Fundamental Catalogue (FK5) reference system requires
specifying an equinox to determine the reference frame.
If required, these additional frame attributes must be specified along with the
coordinate data when a \texttt{Frame} object is created.
\figurename~\ref{fig:frame-transform-graph} shows the network of possible
reference frame transformations as currently implemented in
\texttt{astropy.coordinates}.
Custom user-implemented \texttt{Frame} classes that define transformations to
any reference frame in this graph can then be transformed to any of the other
connected frames.

The typical user does not usually have to interact with the \texttt{Frame} or
\texttt{Representation} classes directly.
Instead, \texttt{astropy.coordinates} provides a high-level interface to
representing astronomical coordinates through the \texttt{SkyCoord} class,
which was designed to provide a single class that
accepts a wide range of possible inputs.
It supports coordinate data in any coordinate frame in any representation by
internally using the \texttt{Frame} and \texttt{Representation} classes.

In what follows, we briefly highlight key new features in
\texttt{astropy.coordinates}.

\subsubsection{Local Earth coordinate frames}
In addition to representing celestial
    coordinates, \astropypkg now supports specifying positions on the Earth in
    a number of different geocentric systems with the \texttt{EarthLocation}
    class.
    With this, \astropypkg now supports Earth-location-specific coordinate
    systems such as the altitude-azimuth (\texttt{AltAz}) or horizontal system.
    Transformations between \texttt{AltAz} and any Barycentric coordinate frame
    also requires specifying a time using the \texttt{Time} class from
    \texttt{astropy.time}.
    With this new functionality, many of the common tasks associated with
    observation planning can now be completed with \astropypkg or the
    \astropy-affiliated package \package{astroplan}\citep{astroplan_AAS}.

\subsubsection{Proper motion and velocity transformations}
    In addition to positional coordinate data, the \texttt{Frame} classes now
    also support velocity data.
    As the default representation for most frames is spherical, most of the
    \texttt{Frame} classes expect proper motion and radial velocity components
    to specify the velocity information.
    The names of the proper motion components all start with \texttt{pm} and
    adopt the same longitude and latitude names as the positional components.
    Transforming coordinates with velocity data is also supported, but in some
    cases the transformed velocity components have limited accuracy because the
    transformations are done numerically instead of analytically.
    The low-level interface for specifying and transforming velocity data  is
    currently experimental.  As such, in version 2.0, only the \texttt{Frame}
    classes (and not the \texttt{SkyCoord} class) support handling velocities.

\subsubsection{Solar System Ephemerides}
    Also new is support for computing ephemerides of major solar system bodies
    and outputting the resulting positions as coordinate objects.
    These ephemerides can be computed either using analytic approximations from
    ERFA or from downloaded JPL ephemerides (the latter requires the
    \package{jplephem}\footnote{\url{https://github.com/brandon-rhodes/python-jplephem}}
    optional dependency and an internet connection).

\subsubsection{Accuracy of coordinate transformations}

In order to check the accuracy of the coordinate transformations in
\package{astropy.coordinates}, we have created a set of benchmarks that we use
to compare transformations between a set of coordinate frames for a number of
packages\footnote{\url{http://www.astropy.org/coordinates-benchmark/summary.html}}.
Since no package can be guaranteed to implement all transformations to
arbitrary precision and some transformations are sometimes subject to
interpretation of standards (in particular in the case of Galactic coordinates),
we do not designate any of the existing packages as the ``ground truth'' but
instead compare each tool to all other tools. The benchmarks are thus useful
beyond the \astropy project since they allow all of the tools to be compared to
all other tools. The tools included in the benchmark at the moment include the
\astropypkg core package, Kapteyn \citep{kapteyn}, NOVAS \citep{novas},
PALpy \citep{pal}, PyAST \citep[a wrapper for AST, described in][]{ast},
PyTPM\footnote{\url{https://github.com/phn/pytpm}}, PyEphem \citep{pyephem},
and pySLALIB \citep[a Python wrapper for SLALIB, described in][]{slalib}.

The benchmarks are meant to evolve over time and include an increasing variety
of cases. At the moment, the benchmarks are set up as follows --
we have generated a standard set of 1000 pairs of random longitudes/latitudes
that we use in all benchmarks. Each benchmark is then defined using an input
and output coordinate frame, using all combinations of FK4, FK5, Galactic,
ICRS, and Ecliptic frames. For now, we set the epoch of observation to J2000.
We also set the frame to J2000 (for FK5 and Ecliptic) and B1950 (for FK4).
In the future, we plan to include a larger variety of epochs and equinoxes,
as well as tests of conversion to/from Altitude/Azimuth. For each benchmark,
we convert the 1000 longitudes/latitudes from the input/output frame with all
tools and quantify the comparison by looking at the median, mean, maximum,
and standard deviation of the absolute separation of the output coordinates
from each pair of tools.

\begin{figure}
  \includegraphics[width=\textwidth]{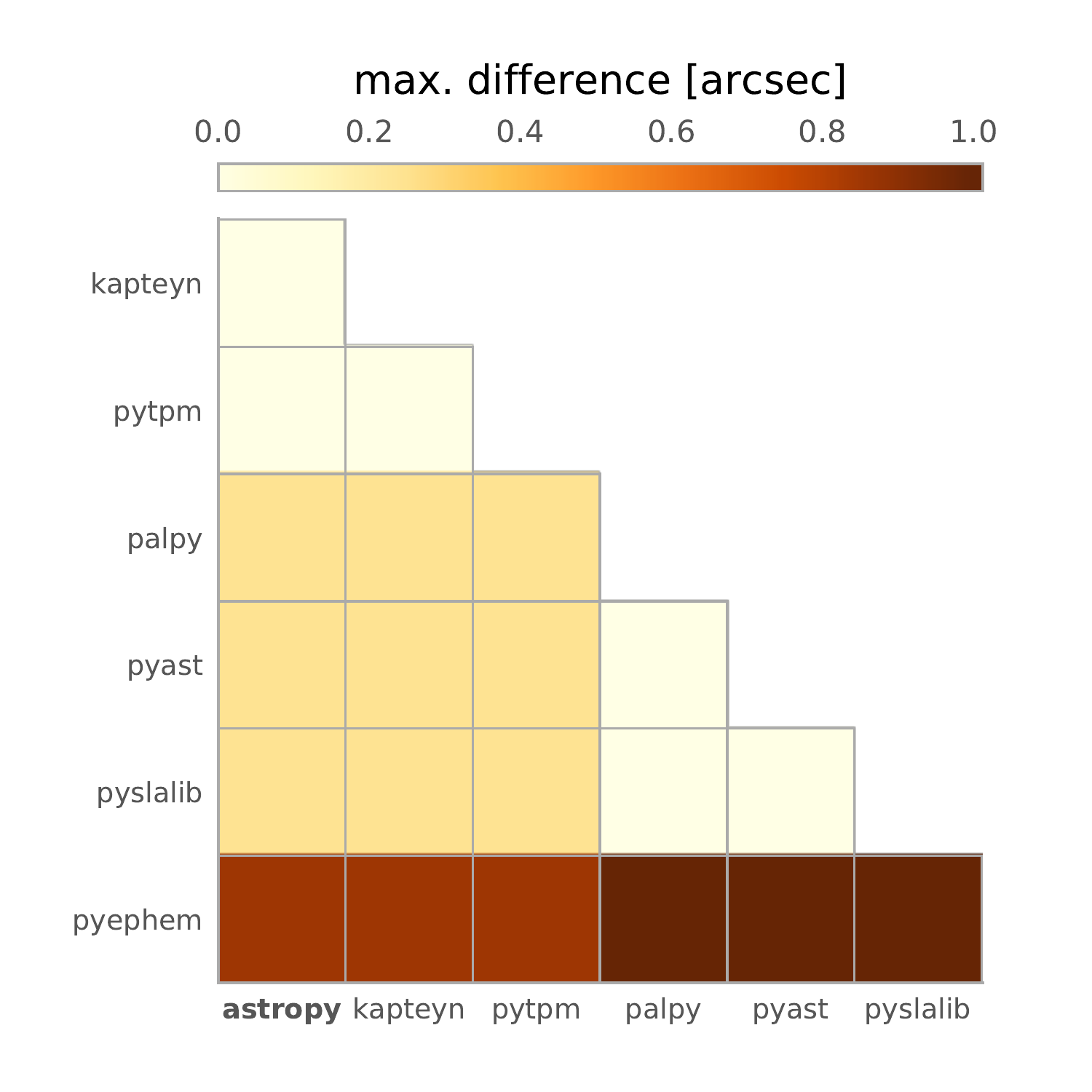}
  \caption{
    Comparison matrix of the maximum difference between longitude-latitude
    values in a set of 1000 random points transformed from FK4 to Galactic
    with the different packages.
    Darker colors (larger differences) are more significant disagreements.
    \label{fig:coordinate_benchmarks}
  }
\end{figure}

\figurename~\ref{fig:coordinate_benchmarks} visualizes the relative accuracy of
the conversion from FK4 to Galactic coordinates for all pairs of tools that
implement this transformation.
In this figure, the color of the cell indicates the maximum difference (in
arcseconds) between the two tools over the 1000 longitude-latitude pairs tested.
This figure shows, for example, that \astropypkg, Kapteyn, and PyTPM agree with
sub-milliarcsecond differences (light colors, small differences), while PALpy,
pySLALIB, and PyAST also agree amongst themselves.
However, there is an offset of around 0.2\arcsec\space between the two groups.
Finally, PyEphem disagrees with all other packages by 0.4--0.8\arcsec (darker
colors, large differences).
These values are only meant to be illustrative and will change over time as the
benchmarks are refined and the packages updated.

\subsection{Time}
\label{sec:time}

The \package{astropy.time} subpackage focuses on supporting time scales (e.g.,
UTC, TAI, UT1) and time formats (e.g., Julian date, modified Julian date) that
are commonly used in astronomy.
This functionality is needed, for example, to calculate barycentric corrections
or sidereal times.
\package{astropy.time} is currently built on the ERFA (\citealt{erfa}) C
library, which replicates the Standards of Fundamental Astronomy (SOFA;
\citealt{sofa}) but is licensed under a three-clause BSD license.
The package was described in detail in  \citet{astropy} and has
stayed stable for the last several versions of \astropypkg.
Thus, in what follows, we only highlight significant changes or new features
since the previous \astropy paper.

\subsubsection{Barycentric and Heliocentric corrections}
Detailed eclipse or transit
timing requires accounting for light travel time differences from the
source to the observatory because of the Earth's motion.
It is therefore common to instead convert times to the Solar System
barycenter or heliocenter where the relative timing of photons is
standardized.
With the location of a source on the sky (i.e., a \texttt{SkyCoord}
object), the location of an observatory on Earth (i.e., an
\texttt{EarthLocation} object), and time values as \texttt{Time}
objects, the time corrections to shift to the solar system barycenter or
heliocenter can now be computed with \package{astropy.time} using the
\texttt{light\_travel\_time} method of a \texttt{Time} object.

\subsection{Data containers}

\subsubsection{\package{nddata}}

The \package{astropy.nddata} subpackage provides three types of functionality: an
abstract interface for representing generic arbitrary-dimensional datasets
intended primarily for subclassing by developers of other packages, concrete
classes building on this interface, and utilities for manipulating these kind of
datasets.

The \texttt{NDDataBase} class provides the abstract interface for gridded data
with attributes for accessing metadata, the world coordinate system (WCS),
uncertainty arrays matched to the shape of the data, and other traits.
Building on this interface, the \texttt{NDData} class provides a minimal
working implementation for storing \package{numpy} arrays. These classes serve
as useful base classes for package authors wishing to develop their own classes
for specific use cases and as containers for exchanging gridded data.

The classes \texttt{NDDataRef}, \texttt{NDDataArray}, and \texttt{CCDData}
extend the base storage functionality with options to do basic arithmetic
(addition, subtraction, multiplication, and division), including error
propagation in limited cases, and slicing of the dataset based on grid
coordinates that appropriately handles masking, errors, and units (if present).
Additionally, the \texttt{CCDData} class also provides reading and writing from
and to FITS files and uses data structures from \astropypkg, like \texttt{WCS},
to represent the file contents abstractly.

The \package{astropy.nddata.utils} module provides utilities that can operate
on either plain \package{numpy} arrays or any of the classes in the
\package{astropy.nddata} subpackage. It features a class for representing
two-dimensional image cutouts, allowing one to easily link pixels in the cutout
to those in the original image or vice versa, to convert
between world and pixel coordinates in the cutout, and to overlay the cutout
on images. Functions to enlarge or reduce an image by doing block replication
or reduction are also provided.

\subsubsection{Tables}
\label{sec:table}

The \package{astropy.table} subpackage provides functionality for
representing and manipulating heterogeneous data. In some respects,
this is similar to \package{numpy} record arrays \citep{numpy} or
\package{pandas} dataframes \citep{pandas} but with modifications for
astronomical data. Most notably, tables from \package{astropy.table}
allow for table or column metadata and can handle
vectors or arrays as table entries.
The subpackage was described in
detail in \cite{astropy}.  Thus, in what follows, we only summarize
key new features or updates to \package{astropy.table} since the
previous \astropy paper. These are support for grouped table
operations, table concatenation, and using array-valued
\package{astropy} objects as table columns.

A table can contain data that naturally form groups; for example, it may
contain multiple observations of a few sources at different points in time
and in different bands. Then, we may want to split the table into groups based
on the combination of source observed and the band, after which we combine the
results for each combination of source and band in some way (e.g., finding
the mean or standard deviation of the fluxes or magnitudes over time) or filter
the groups based on user-defined criteria. These kinds of grouping and
aggregation operations are now fully supported by \texttt{Table} objects.

\texttt{Table} objects can now be combined in several different ways. If two
tables have the same columns, we may want to stack them ``vertically'' to create a
new table with the same columns but all rows. If two tables are row-matched but
have distinct columns, we may want to stack them ``horizontally'' to create a
new table with the same rows but all columns. For other situations, more generic
table concatenation or join are also possible when two tables share some
columns.

The \texttt{Table} object now allows array-valued \texttt{Quantity}, celestial
coordinate (\texttt{SkyCoord}), and date/time (\texttt{Time}) objects to
be used as columns. It also provides a general way for other user-defined
array-like objects to be used as columns.
This makes it possible, for instance, to easily
represent catalogs of sources or time series in \astropy, while having both the
benefits of the \texttt{Table} object (e.g., accessing specific rows/columns
or groups of them and combining tables) and of, for example,
the \texttt{SkyCoord} or the \texttt{Time} classes (e.g., converting the
coordinates to a different frame or accessing the date/time in the desired time scale).

\subsection{\package{io}}

The \package{astropy.io} subpackage provides support for reading and writing
data to a variety of ASCII and binary file formats, such as a wide range of
ASCII data table formats, FITS, HDF5, and VOTable.
It also provides a unified interface for reading and writing data with these
different formats using the \package{astropy.table} subpackage.
For many common cases, this simplifies the process of file input and output (I/O) and
reduces the need to master the separate details of all the I/O packages within
\astropypkg. The file interface allows transparent compression of
the \texttt{gzip}, \texttt{bzip2} and \texttt{lzma} (\texttt{.xz})
formats; for the latter two if the Python installation was compiled
with support the respective libraries.

\subsubsection{ASCII}

One of the problems when storing a table in an ASCII format is
preserving table meta-data such as comments, keywords and column data
types, units, and descriptions. The newly defined \emph{Enhanced
Character Separated Values} (ECSV,  \citealt{ape6}) format makes it
possible to write a table to an ASCII-format file and read it back
with no loss of information. The ECSV format has been designed to be
both human-readable and compatible with most simple CSV readers.

The \package{astropy.io.ascii} subpackage now includes a significantly faster
Cython/C engine for reading and writing ASCII files. This is available for most
of the common formats. It also offers some additional features like parsing
of different exponential notation styles, such as commonly produced
by Fortran programs.  On average, the new engine is about 4 to 5 times faster
than the corresponding pure-\python implementation and is often comparable to
the speed of the \package{pandas} \citep{pandas} ASCII file interface.  The
fast reader has a parallel processing option that allows harnessing multiple
cores for input parsing to achieve even greater speed gains.  By default,
\texttt{read()} and \texttt{write()} will attempt to use the fast Cython/C engine
when dealing with compatible formats. Certain features of the full read
/ write interface are unavailable in the fast version, in which case the
reader will by default fall back automatically to the pure-\python version.

The \package{astropy.io.ascii} subpackage now provides the capability
to read a table within an HTML file or web URL into an \astropypkg
\texttt{Table} object. A \texttt{Table} object can now also
be written out as an HTML table.

\subsubsection{FITS}

The \package{astropy.io.fits} subpackage started as a direct port of the
PyFITS project \citep{PyFITS}. Therefore, it is pretty stable, with mostly bug
fixes but also a few new features and performance improvements.  The API
remains mostly compatible with PyFITS, which is now deprecated in favor of
\astropypkg.

Command-line scripts are now available for printing a summary of the HDUs in
FITS file(s) (\texttt{fitsinfo}) and for printing the header information to
the screen in a human-readable format (\texttt{fitsheader}).

FITS files are now loaded \emph{lazily} by default, i.e., an object
representing the list of HDUs is created but the data are not loaded into
memory until requested.  This approach should provide substantial speed-ups
when using the convenience functions (e.g., \texttt{getheader()} or
\texttt{getdata()}) to get an HDU that is near the beginning in a file with
many HDUs.

\subsection{Modeling}
\label{sec:modeling}

The \package{astropy.modeling} subpackage provides a framework for representing
analytical models and performing model evaluation and parameter fitting.
The main motivation for this functionality was to create a framework that
allows arbitrary combination of models to support the Generalized World
Coordinate System (GWCS) package.\footnote{\url{https://github.com/spacetelescope/gwcs}}
The current FITS WCS specification lacks the flexibility to represent
arbitrary distortions and does not meet the needs of many types of current instrumentation.
The fact that the \astropypkg modeling framework now supports propagating
units also makes it a useful tool for representing and fitting astrophysical
models within data analysis tools.

Models and fitters are independent of each other: a model can be fit with different
fitters and new fitters can be added without changing existing models. The
framework is designed to be flexible and easily extensible. The goal is to have
a rich set of models, but to also facilitate creating new ones, if necessary.

\subsubsection{Single Model Definition and Evaluation}

Models are defined by their parameters and initialized by providing values for them. The
names of the parameters are stored in a list, \texttt{Model.param\_names}. Parameters are complex objects.
They store additional information -- default value, default unit, and parameter constraints. Parameter
values and constraints can be updated by assignment. Supported constraints include \texttt{fixed},
and \texttt{tied} parameters, and \texttt{bounds} on parameter values. The framework also supports models
for which the number of parameters and their names are defined by another argument.
A typical example is a polynomial model defined by its degree.
A model is evaluated by calling it as a function.

If an analytical inverse of a model exists it can be
accessed  by calling \texttt{Model.inverse}. In addition, \texttt{Model.inverse} can be assigned
another model which represents a computed inverse.

Another useful settable property of models is \texttt{Model.bounding\_box}. This attribute sets the domain over which the model is defined. This greatly improves the efficiency of evaluation when the input range is much larger than the characteristic width of the model itself.

\subsubsection{Model Sets}

\package{astropy.modeling} provides an efficient way to set up the same type of model with many different sets of parameter values.
This creates a model set that can be efficiently evaluated. For example, in PSF (point spread function) photometry, all objects in an image will have a PSF of the same functional form, but with different positions and amplitudes.

\subsubsection{Compound Models}
Models can be combined using arithmetic expressions. The result is also a model, which can further be combined with other models. Modeling supports arithmetic (+, -, *, /, and **), join ($\&$), and composition ($|$) operators. The rules for combining models involve matching their inputs and outputs. For example, the composition operator, $|$, requires the number of outputs of the left model to be equal to the number of inputs of the right one. For the join operator, the total number of inputs must equal the sum of number of inputs of both the left and the right models. For all arithmetic operators, the left and the right models must have the same number of inputs and outputs. An example of a compound model could be a spectrum with interstellar absorption. The stellar spectrum and the interstellar extinction are represented by separate models, but the observed spectrum is fitted with a compound model that combines both.

\subsubsection{Fitting Models to Data}

\package{astropy.modeling} provides several fitters which are wrappers around some of the \texttt{numpy} and \texttt{scipy.optimize} functions and provide support for specifying parameter constraints. The fitters take a model and data as input and return a copy of the model with the optimized parameter values set. The goal is to make it easy to extend the fitting framework to create new fitters. The optimizers available in \astropypkg version 2.0 are Levenberg--Marquardt (\texttt{scipy.optimize.leastsq}), Simplex (\texttt{scipy.optimize.fmin}), SLSQP (\texttt{scipy.optimize.slsqp}), and LinearLSQFitter (\texttt{numpy.linalg.lstsq} which provides exact solutions for linear models).

Modeling also supports a plugin system for fitters, which allows using the
\astropypkg models with external fitters. An example of this is
\package{SABA}\footnote{\url{https://github.com/astropy/saba}}, which is a bridge between
Sherpa \citep{sherpa},
and \package{astropy.modeling}, to bring the Sherpa fitters into \astropypkg.

\subsubsection{Creating New Models}
If arithmetic combinations of existing models is not sufficient, new model
classes can be defined in different ways. The \package{astropy.modeling}
package provides tools to turn a simple function into a full-featured model,
but it also allows extending the built-in model class with arbitrary code.

\subsubsection{Unit Support}

The \package{astropy.modeling} subpackage now supports the representation,
evaluation, and fitting of models using \texttt{Quantity} objects, which attach
units to scalar values or arrays of values. In practice, this means that one
can, for example, fit a model to data with units and get parameters that also
have units out, or initialize a model with parameters with units and evaluate
it using input values with different but equivalent units. For example, the
blackbody model (\texttt{BlackBody1D}) can be used to fit observed flux
densities in a variety of units and as a function of different units of
spectral coordinates (e.g., wavelength or frequency).

\subsection{Convolution}

\begin{figure}
\includegraphics[width=\textwidth]{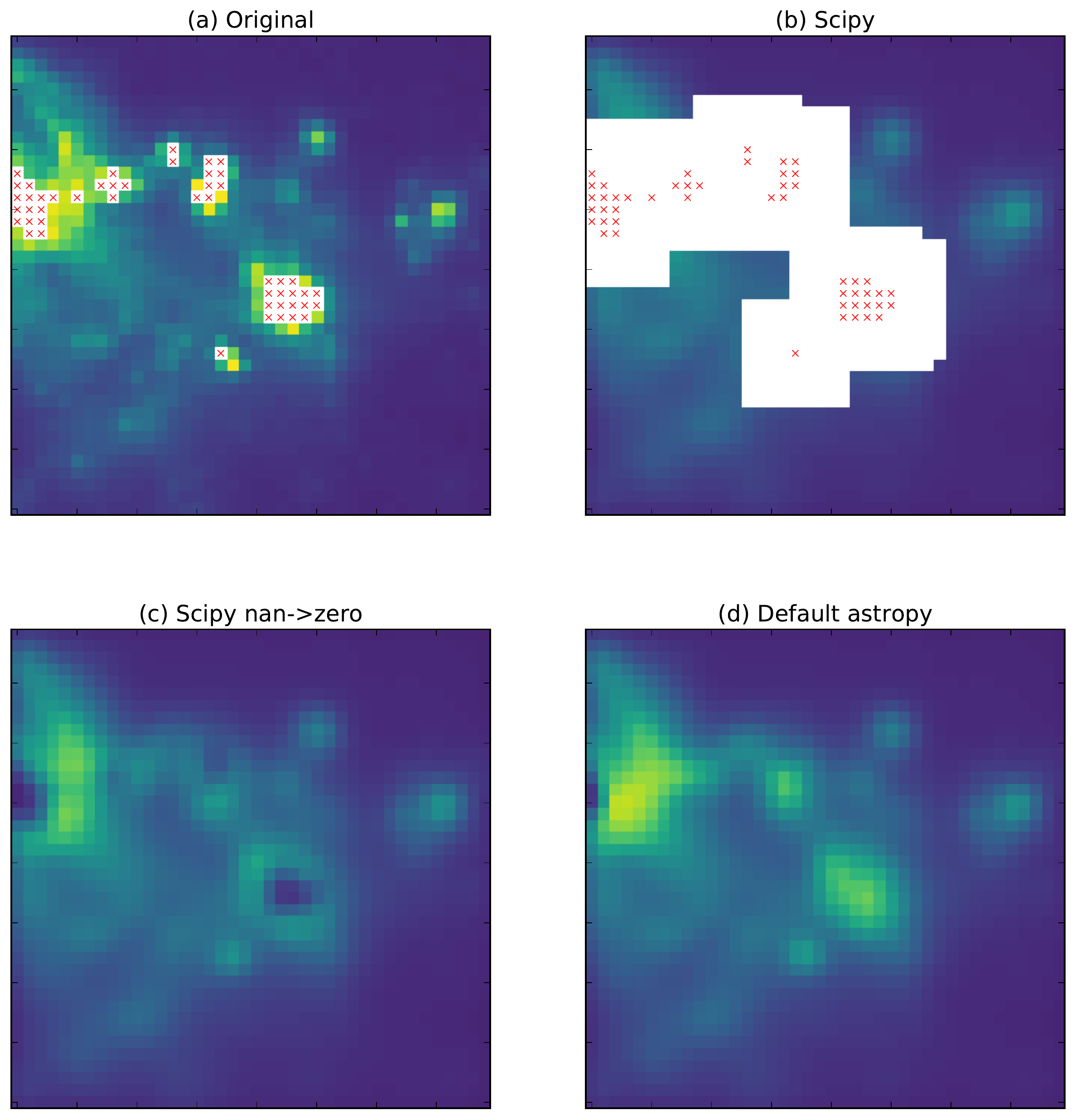}
\caption{%
    An example showing different modes of convolution available in the \python
    ecosystem.  Each red $x$ signifies a pixel that is set to \texttt{NaN} in the
    original data (a).  If the data are convolved with a Gaussian kernel on a
    $9\times 9$ grid using
    \package{scipy}'s direct convolution (b), any pixel within range of the original \texttt{NaN}
    pixels is also set to \texttt{NaN}.  Panel (c) shows what happens if the
    \texttt{NaN}s are set to zero first: the originally \texttt{NaN} regions are
    depressed relative to their surroundings.  Finally, panel (d) shows
    \astropypkg's convolution behavior, where the missing pixels are replaced
    with values interpolated from their surroundings using the convolution
    kernel.
    \label{fig:convolution-example}
}
\end{figure}

The \package{astropy.convolution} subpackage implements
\textit{normalized convolution} \citep[e.g.,][]{Knutsson1993}, an image
reconstruction technique in which missing data are ignored during the
convolution and replaced with values interpolated using the kernel.
An example is given in \figurename~\ref{fig:convolution-example}.
In \astropypkg versions $\leq 1.3$, the direct convolution and Fast Fourier Transform (FFT)
convolution approaches were inconsistent, with only the latter implementing
normalized convolution.
As of version 2.0, the two methods now agree and include a suite of
consistency checks.

\subsection{Visualization}

The \package{astropy.visualization} subpackage provides functionality that can
be helpful when visualizing data. This includes a framework (previously the
standalone \package{wcsaxes} package) for plotting
astronomical images with coordinates with \package{matplotlib}, functionality related to image
normalization (including both scaling and stretching), smart histogram
plotting, red-green-blue (RGB) color image creation from separate images,
and custom plotting styles for \package{matplotlib}.

\subsubsection{Image Stretching and Normalization}

\label{sec:stretch}

\package{astropy.visualization} provides a framework for transforming values in
images (and more generally any arrays), typically for the purpose of
visualization. Two main types of transformations are normalization and
stretching of image values.

Normalization transforms the image values to the range $[0,1]$ using lower and
upper limits $(v_{\rm min}, v_{\rm max})$,
\begin{equation}
y = \frac{x - v_{\rm min}}{v_{\rm max} - v_{\rm min}} \quad ,
\end{equation}
where $x$ represents the values in the original image.

Stretching transforms the image values in the range $[0,1]$ again to the range
$[0,1]$ using a linear or non-linear function,
\begin{equation}
z = f(y) \quad .
\end{equation}

Several classes are provided for automatically determining intervals
(e.g., using image percentiles) and for normalizing values in this interval to
the $[0,1]$ range.

\package{matplotlib} allows a custom normalization and stretch to be used when
displaying data by passing in a normalization object.
The \package{astropy.visualization} package also provides a normalization class
that wraps the interval and stretches objects into a normalization object that
\package{matplotlib} understands.

\subsubsection{Plotting image data with world coordinates}

Astronomers dealing with observational imaging commonly need to make figures
with images that include the correct coordinates and optionally display a
coordinate grid. The challenge, however, is that the conceptual coordinate axes
(such as longitude/latitude) need not be lined up with the pixel axes of the
image. The \package{astropy.visualization.wcsaxes} subpackage implements a
generalized way of making figures from an image array and a WCS object
that provides the transformation between pixel and world coordinates.

World coordinates can be, for example, right ascension and declination, but can
also include, for example, velocity, wavelength, frequency, or time.
The main features from this subpackage include the ability to control which
axes to show which coordinate on (e.g., showing longitude ticks on the
top and bottom axes and latitude on the left and right axes), controlling the
spacing of the ticks either by specifying the positions to use or providing a
tick spacing or an average number of ticks that should be present on each axis,
setting the format for the tick labels to ones commonly used by astronomers,
controlling the visibility of the grid/graticule, and overlaying ticks, labels,
and/or grid lines from different coordinate systems. In addition, it is
possible to pass data with more than two dimensions and slice on-the-fly.
Last but not least, it is also able to define non-rectangular frames, such as,
for example, Aitoff projections.

This subpackage differs from \package{APLpy} \citep{aplpy}, in that the latter
focuses on providing a very high-level interface to plotting that requires very
few lines of code to get a good result, whereas \package{wcsaxes} defines an
interface that is much closer to that of \package{matplotlib} \citep{matplotlib}.
This enables significantly more advanced visualizations.

\begin{figure}
\includegraphics[width=\textwidth]{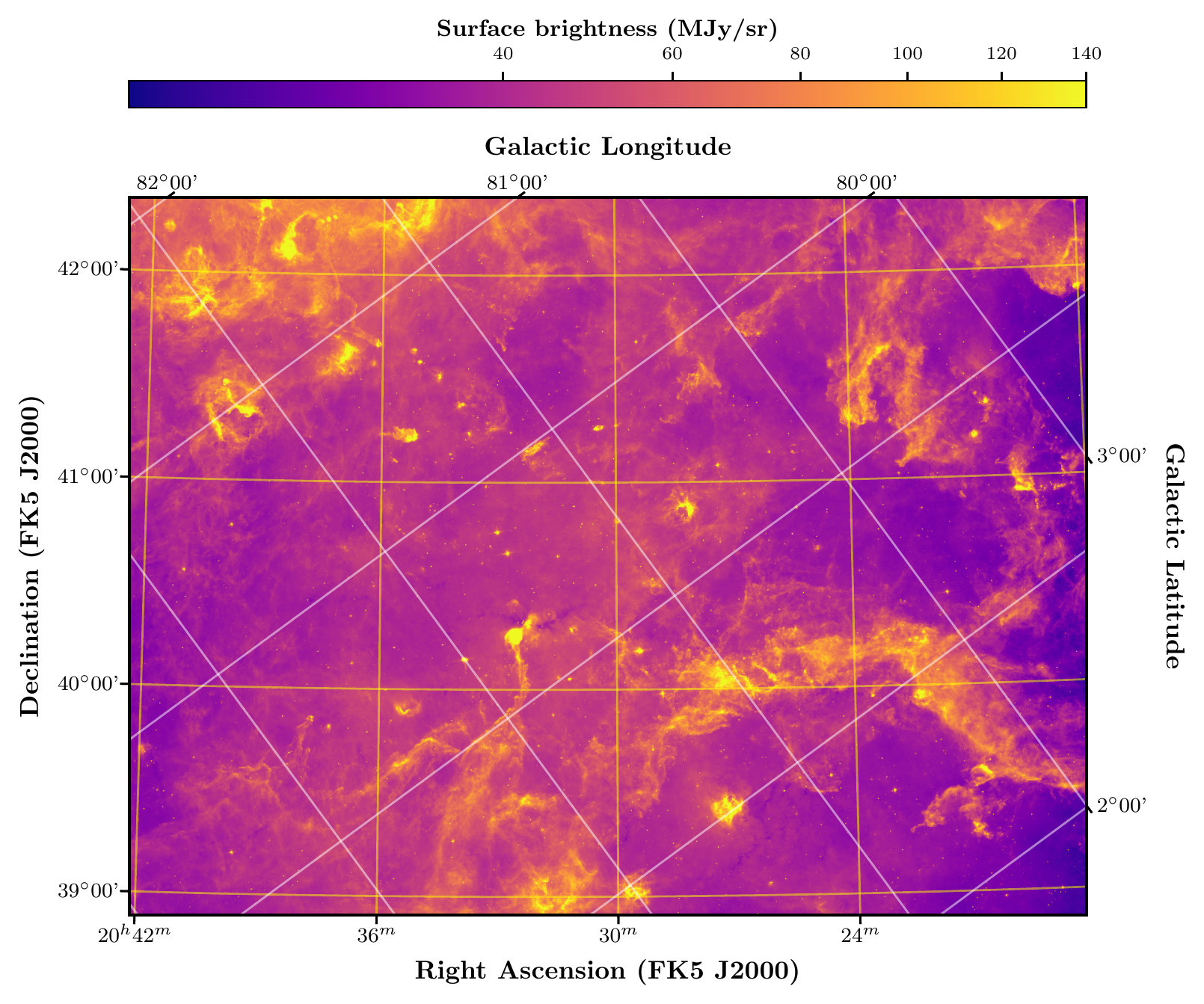}
\caption{%
An example of figure made using the \package{astropy.visualization.wcsaxes}
subpackage, using \textit{Spitzer}/IRAC 8.0~$\mu$m data from the Cygnus-X
\textit{Spitzer} Legacy survey \citep{cygnusx}.
\label{fig:wcsaxes}
}
\end{figure}

An example of a visualization made with \package{wcsaxes} is shown in
\figurename~\ref{fig:wcsaxes}. This example illustrates the ability to
overlay multiple coordinate systems and customize which ticks/labels are shown
on which axes around the image. This also uses the image stretching
functionality from \sectionname~\ref{sec:stretch} to show the image in a
square-root stretch (automatically updating the tick positions in the colorbar).

\subsubsection{Choosing Histogram Bins}

\package{astropy.visualization} also provides a histogram function, which is a
generalization of \texttt{matplotlib}’s histogram function, to allow for a more
flexible specification of histogram bins.  The function provides several methods
of automatically tuning the histogram bin size. It has a syntax identical to
\texttt{matplotlib}’s histogram function, with the exception of the \texttt{bins}
parameter, which allows specification of one of four different methods for
automatic bin selection: ``blocks'', ``knuth'', ``scott'', or ``freedman''.

\subsubsection{Creating color RGB images}

\cite{Lupton2004} describe an ``optimal'' algorithm for producing
RGB composite images from three separate high-dynamic range
arrays. The \package{astropy.visualization} subpackage provides a convenience
function to create such a color image.  It also includes an associated set of
classes to provide alternate scalings.
This functionality was contributed by developers from the Large Synoptic Survey
Telescope (LSST) and serves as an example of contribution to \astropy from a
more traditional engineering organization \citep{Jennes2016}.

\begin{figure}
\includegraphics[width=\textwidth]{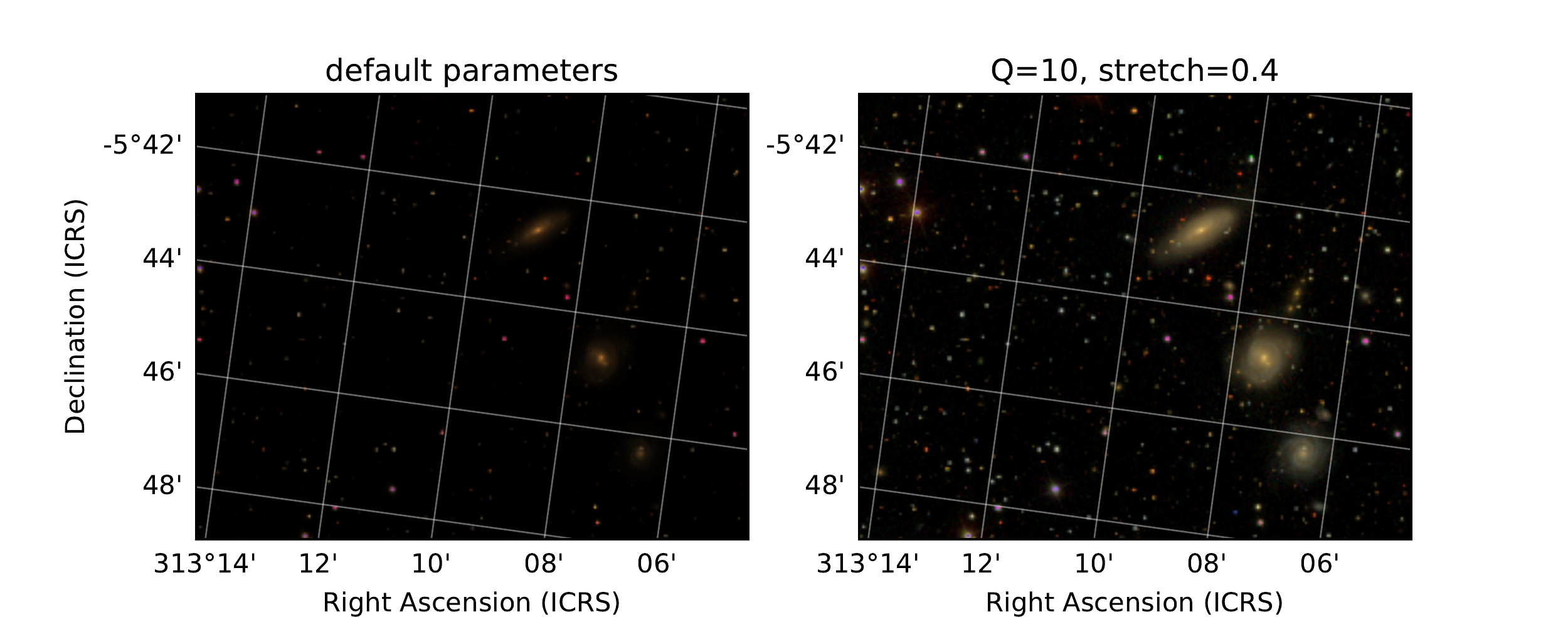}
\caption{An RGB color image of the region near the Hickson 88 group
constructed from SDSS images and the \package{astropy.visualization}
tools.
This example uses \package{astropy.visualization.wcsaxes} to display the
sky coordinate grid, and the \texttt{astropy.visualization.make\_lupton\_rgb()}
function to produce the RGB image from three SDSS filter images ($g$, $r$, $i$).
The left and right panel images show two different parameter choices for the
stretch and softening parameters (shown in the titles).
\label{fig:ngc6977}}
\end{figure}

The Sloan Digital Sky Survey (SDSS) SkyServer color images were made using a
variation on this technique.  As an example, in \figurename~\ref{fig:ngc6977},
we show an RGB color image of the Hickson 88 group, centered near
NGC~6977.\footnote{\url{http://skyserver.sdss.org/dr13/en/tools/chart/navi.aspx?ra=313.12381&dec=-5.74611}}
This image was generated from SDSS images using the
\package{astropy.visualization} tools.

\subsection{Cosmology}

The \package{cosmology} subpackage contains classes for representing different
cosmologies and functions for calculating commonly used quantities such as
look-back time and distance.   The subpackage was described in detail in
\cite{astropy}.  The default cosmology in \astropypkg version 2.0 is given by
the values in \cite{2016A&A...594A..13P}.

\subsection{Statistics}

The \package{astropy.stats} package provides statistical tools that
are useful for astronomy and are either not found in or extend
the available functionality of other \python statistics packages, such
as \package{scipy} \citep{scipy} or \package{statsmodels}
\citep{seabold2010statsmodels}.  \package{astropy.stats} contains
a range of functionality used by many different disciplines
in astronomy. It is not a complete set of statistical tools, but rather
a still growing collection of useful features.

\subsubsection{Robust Statistical Estimators}

Robust statistics provide reliable estimates of basic statistics for complex
distributions that largely mitigate the effects of outliers.
\package{astropy.stats} includes several robust statistical functions that are
commonly used in astronomy, such as sigma clipping methods for rejecting
outliers, median absolute deviation functions, and biweight estimators,
which have been used to calculate the velocity dispersion of galaxy clusters
\citep{Beers1990}.

\subsubsection{Circular Statistics}

Astronomers often need to compute statistics of quantities evaluated on a
circle, such as sky direction or polarization angle.
A set of circular statistical estimators based on \citet{JammalamadakaSengupta}
are implemented in \package{astropy.stats}.  These functions provide
measurements of the circular mean, variance, and moment.   All of these
functions work with both \texttt{numpy.ndarrays} (assumed to be in radians) and
\texttt{Quantity} objects.  In addition, the subpackage includes
tests for Rayleigh Test, \texttt{vtest}, and a function to compute the maximum
likelihood estimator for the parameters of the von Mises distribution.

\subsubsection{Lomb-Scargle Periodograms}

Periodic analysis of unevenly-spaced time series is common across many
sub-fields of astronomy. The \package{astropy.stats} package now includes
several efficient implementations of the Lomb-Scargle periodogram
\citep{Lomb76, Scargle82} and several generalizations, including floating mean
models \citep{Zechmeister09}, truncated Fourier models \citep{Bretthorst2003},
and appropriate handling of heteroscedastic uncertainties.
Importantly, the implementations make use of several fast and scalable
computational approaches \citep[e.g.,][]{Press89, Palmer09}, and thus can be
applied to much larger datasets than Lomb-Scargle algorithms available in,
e.g., \package{scipy.stats} (\citealt{scipy}). Much of the Lomb-Scargle code
in \astropypkg has been adapted from previously-published open-source code
\citep{astroML, VanderPlas2015}.

\subsubsection{Bayesian Blocks and Histogram Binning}
\begin{figure}
\includegraphics[width=\textwidth]{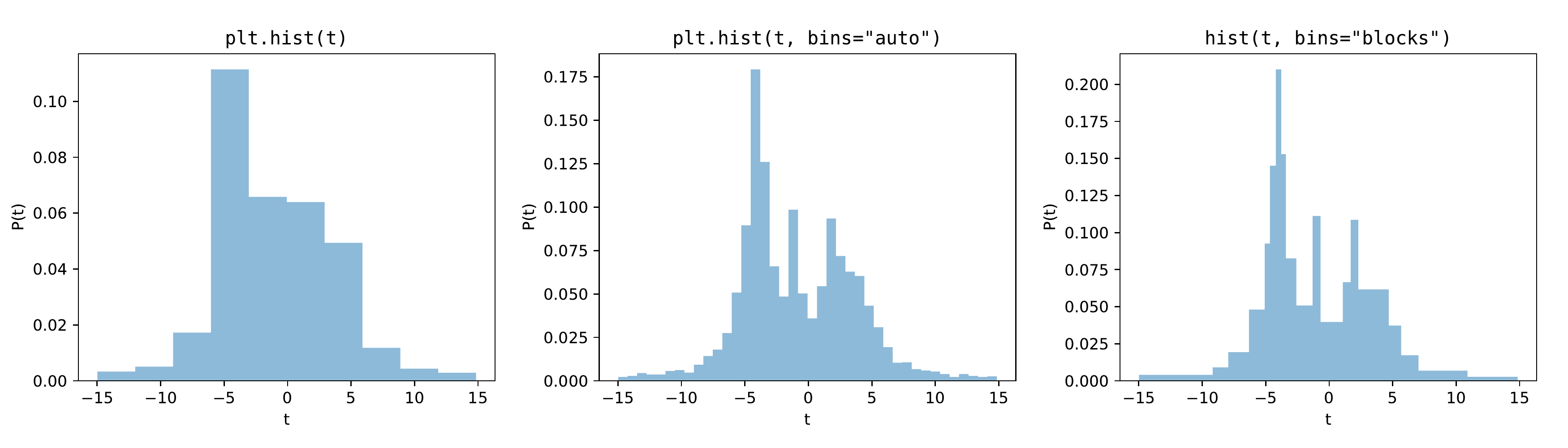}
\caption{%
    Three approaches to a 1D histogram:
    {\it left:} a standard histogram using \package{matplotlib}'s default of 10 bins.
    {\it center:} a histogram with the number of equal-width bins determined automatically using \package{numpy}'s {\tt bins='auto'}.
    {\it right:} a histogram created with \astropypkg, with irregularly-spaced bins computed via the Bayesian Blocks algorithm.
    Compared to regularly-spaced bins, the irregular bin widths give a more accurate visual representation of features in the dataset at various scales.
    \label{fig:bayes-blocks-hist}
}
\end{figure}

\package{astropy.stats} also includes an implementation of
{\it Bayesian Blocks} \citep{Scargle2013}, an algorithm for analysis of
break-points in non-periodic astronomical time-series. One interesting
application of Bayesian Blocks is its use in determining optimal histogram
binnings, particularly binnings with unequal bin sizes.
This code was adapted, with several improvements, from the \package{astroML}
package \citep{astroML}. An example of a histogram fit using the Bayesian
Blocks algorithm is shown in the right panel of
\figurename~\ref{fig:bayes-blocks-hist}.

\section{Infrastructure for Astropy affiliated packages}

\label{sec:infrastructure}

In addition to astronomy-specific packages and libraries, the \astropy Project
also maintains and distributes several general-purpose infrastructure packages
that assist the maintenance and upkeep of the \astropypkg core package and
other affiliated packages.
The following sections describe the most widely-used infrastructure packages
developed by the \astropy Project.

\subsection{Package template}

\astropy provides a package template --- as a separate \github repository,
\package{astropy/package-template}\footnote{\url{https://github.com/astropy/package-template/}}
--- that aims to simplify setting up packaging, testing, and
documentation builds for developers of affiliated packages or
\astropypkg-dependent packages.
Any \python package can make use of this ready-to-go package layout, setup,
installation, and \package{Sphinx} documentation build infrastructure that was
originally developed for the \astropypkg core package and affiliated packages
maintained by the \astropy project.
The package template also provides a testing framework, template configurations
for continuous integration services, and \package{Cython} build support.

\subsection{Continuous integration helpers}

\astropy also provides a set of scripts for setting up and configuring
continuous integration (CI) services as a \github repository,
\package{astropy/ci-helpers}.\footnote{\url{https://github.com/astropy/ci-helpers}}
These tools aim to enable package maintainers to control their testing setup
and installation process for various CI services through a
set of environment variables.
While the current development is mostly driven by the needs of the \astropy
ecosystem, the actual usage of this package is extremely widespread. The current
tools support configuration for Travis CI\footnote{\url{https://travis-ci.org/}} and
Appveyor CI\footnote{\url{https://www.appveyor.com/}}.

\subsection{Sphinx extensions}

The documentation for many \python packages, including
all the packages in the \astropy ecosystem, is written using the
\package{Sphinx} documentation build system.
\package{Sphinx} supports writing documentation using plain text files
that follow a markup language called \texttt{reStructuredText} (RST).
These files are then transformed into HTML, PDF, or \LaTeX{} documents
during the documentation build process.
For the \astropy project, we have developed several \package{Sphinx} extensions
that facilitate automatically generating API documentation for large projects,
like the \astropypkg core package.
The main extension we have developed is
\package{sphinx-automodapi}\footnote{\url{http://sphinx-automodapi.readthedocs.io}},
which provides a convenient single RST command to generate a set of
documentation pages, listing all of the available classes, functions, and
attributes in a given \python module.

\section{The future of the Astropy project}
\label{sec:future}

Following the release of version 2.0, development on the next major version of
the \astropypkg core package (version 3.0) has already begun.
On top of planned changes and additions to the core package, we also plan to
overhaul the \astropy educational/learning materials and further
generalize the infrastructure utilities originally developed for the core
package for the benefit of the community.

\subsection{Future versions of the Astropy core and affiliated packages}

One of the most significant changes coming in this next major release will be
removing the support for \python 2 \citep{ape10}: future versions of \astropypkg
will only support \python 3.5 or higher.
Removing \python 2 support will allow the use of new \python 3-only features,
simplify the code base, and reduce the testing overhead for the package.
\astropypkg version 3.0 is currently scheduled for January 2018.

In the next major release after version 3.0, scheduled for
mid-2018, the focus will be on algorithm optimization and
documentation improvement.
To prepare for this release, we are subjecting the core package to testing,
evaluation, and performance monitoring.
As a result, less new functionality may be introduced as a trade-off for
better performance.

Beyond the core code, the \astropy project is also further developing the
\astropy-managed affiliated packages.
While these may not be integrated into the \astropypkg core package, these
projects provide code that is useful to the astronomical community and meet the
testing and documentation standards of \astropy.
Some of these new efforts include an initiative to develop tools for
spectroscopy \citep[\package{specutils}, \package{specreduc}, \package{specviz}]
{ape13}, integration of LSST software, and support for HEALPIX
projection. %\citep{astropy-healpix}
%\inlinecomment{BMS}{TODO: uncomment references of packages once they are available}

\subsection{Learn Astropy}

The documentation of the \astropypkg core package contains narrative descriptions of
the package's functionality, along with detailed usage notes for functions,
classes, and modules.
While useful as a reference for more experienced \python users, it is not the
proper entry-point for other users or learning environments.
In the near future, we will launch a new resource for learning to use both the
\astropypkg core package and the many packages in the broader \astropy
ecosystem, under the name \emph{Learn Astropy}.

The new \emph{Learn Astropy} site will present several different ways to engage
with the \astropy ecosystem:
\begin{description}
    \item[Documentation] The \astropypkg and affiliated package documentation
        contains the complete description of a package with all requisite
        details, including usage, dependencies, and examples.
        The pages will largely remain as-is, but will be focused towards more
        intermediate users and as a reference resource.
    \item[Examples] These are stand-alone code snippets that live in the
        \astropypkg documentation that demonstrate a specific functionality
        within a subpackage.
        The \astropypkg core package documentation will then gain a new ``index
        of examples'' that links to all of the code or demonstrative examples
        within any documentation page.
    \item[Tutorials] The \astropy tutorials are step-by-step demonstrations of
        common tasks that incorporate several packages or subpackages.
        Tutorials are more extended and comprehensive than examples, may contain
        exercises for the users, and are generally geared towards workshops or
        teaching.
        Several tutorials already
        exist\footnote{\url{http://tutorials.astropy.org/}} and are being
        actively expanded.
    \item[Guides] These are long-form narrative, comprehensive, and
        conceptually-focused documents (roughly one book chapter in length),
        providing stand-alone introductions to core packages in addition to the
        underlying astronomical concepts.
        These are less specific and more conceptual than tutorials.
        For example, ``using \astropypkg and \package{ccdproc} to reduce imaging
        data.''
\end{description}
We encourage any users who wish to see specific material to either contribute or
comment on these efforts via the \astropy mailing list or \package{astropy/astropy-tutorials}
\github repository.\footnote{\url{https://github.com/astropy/astropy-tutorials}}

\section{Conclusion}
\label{sec:conclusion}

The development of the \astropypkg package and cultivation of the \astropy
ecosystem are still maintaining significant growth while improving in stability,
breadth, and reliability.
As the \astropypkg core package becomes more mature, several subpackages have
reached stability with a rich set of features that help astronomers worldwide to
perform many daily tasks, such as planning observations, analyzing data or
simulation results, and writing publications.
The strong emphasis that the \astropy project puts on reliability and high coding
standards helps users to trust the calculations performed with \astropypkg and
to publish reproducible results.
At the same time, the \astropy ecosystem and core package are both growing: new
functionality is still being contributed and new affiliated packages are being
developed to support more specialized needs.

The \astropy project is also spreading awareness of best practices in
community-driven software development.
This is important as most practicing astronomers were not explicitly taught
computer science and software development, despite the fact that a substantial
fraction of many astronomers' workload today is related to software use
and development.
The \astropypkg package leads by example, showing all interested astronomers how
modern tools like \texttt{git} version control or CI testing can
increase the quality, accessibility, and discoverability of astronomical
software without overly complicating the development cycle.
Within \astropy, all submitted code is reviewed by at least one, but typically
more, member of the \astropy community, who provide feedback to contributors,
which helps to improve their software development skills.
As a community, \astropy follows an explicit code of conduct \citep{ape8} and treats all
contributors and users with respect, provides a harassment-free environment, and
encourages and welcomes new contributions from all.
Thus, while the \astropy project provides and develops software and tools
essential to modern astronomical research, it also helps to prepare the current
and next generation of researchers with the knowledge to adequately use,
develop, and contribute to those tools within a conscientious and welcoming
community.

\acknowledgments

We would like to thank the members of the community that have
contributed to \astropy,
that have opened issues and provided feedback, and have supported the
project in a number of different ways.  We would like to acknowledge
Alex Conley and Neil Crighton for maintaining the  \package{cosmology}
subpackage.

The \astropy community is supported by and makes use
of a number of organizations and services outside the traditional
academic community. We thank Google for financing and organizing the
Google Summer of Code (GSoC) program, that has funded severals
students per year to work on \astropy related projects over the
summer. These students often turn into long-term contributors. We also
thank NumFOCUS and the Python Software Foundation for financial
support. Within the academic community, we thank
institutions that make it possible that astronomers and other developers on
their staff can contribute their time to the development of
\astropy projects.  We would like acknowledge the support of the
Space Telescope Science Institute, Harvard–Smithsonian Center for Astrophysics,
and the South African Astronomical Observatory.

The following individuals would like to recognize support for their personal contributions.
HMG was supported by the National Aeronautics and Space Administration through the
Smithsonian Astrophysical Observatory contract SV3-73016 to MIT for Support of the Chandra X-Ray Center, which is
operated by the Smithsonian Astrophysical Observatory for and on behalf of the National Aeronautics Space
Administration under contract NAS8-03060. JTV was supported by the UW eScience Institute, via grants from the Moore
Foundation, the Sloan Foundation, and the Washington Research Foundation. SMC acknowledges the National Research
Foundation of South Africa. TLA was supported by NASA contract NAS8-03060. Support for E.J.T. was provided by NASA
through Hubble Fellowship grant No. 51316.01 awarded by the Space Telescope Science Institute, which is operated by
the Association of Universities for Research in Astronomy, Inc., for NASA, under contract NAS 5-26555, as well as a
Giacconi Fellowship. MB was supported by the FONDECYT regular project 1170618 and the MINEDUC-UA projects codes
ANT 1655 and ANT 1656. DH was supported through the SFB 881 ``The
Milky Way System'' by the German Research Foundation
(DFG). WEK was supported by an ESO Fellowship. C.M. is supported by NSF grant AST-1313484. SP was supported by
grant AYA2016-75808-R (FEDER) issued by the Spanish government. JEHT was supported by the Gemini Observatory, which
is operated by the Association of Universities for Research in Astronomy, Inc., on behalf of the international
Gemini partnership of Argentina, Brazil, Canada, Chile, and the United States of America.

Furthermore, the \astropy packages would not exist
in their current form without a number of web services for code
hosting, continuous integration, and documentation; in particular,
\astropy heavily relies on GitHub, Travis CI, Appveyor, CircleCI, and
Read the Docs.

\astropypkg interfaces with the SIMBAD database,
operated at CDS, Strasbourg, France. It also makes use of the ERFA library \citep{erfa},
which in turn derives from the IAU SOFA Collection\footnote{\url{http://www.iausofa.org}} developed by the International Astronomical Union Standards of Fundamental Astronomy \citep{sofa}.

\software{\package{astropy} (\citealt{astropy}),
          \package{numpy} (\citealt{numpy}),
          \package{scipy} (\citealt{scipy}),
          \package{matplotlib} (\citealt{matplotlib}),
          \package{Cython} (\citealt{cython}),
          \package{SOFA} (\citealt{sofa}),
          \package{ERFA} (\citealt{erfa})
          }

\bibliographystyle{aasjournal}
\bibliography{bibliography}

\appendix

\section{List of Affiliated Packages}

\begin{longrotatetable}
    \begin{deluxetable*}{cccp{3in}c}
    \tablecaption{Registry of affiliated packages.}
    \label{tab:registry}
    \tablehead{
        \colhead{Package Name} &
        \colhead{Stable} &
        \colhead{PyPI Name} &
        \colhead{Maintainer} &
        \colhead{Citation}
      }
      \startdata
        \href{http://github.com/aplpy/aplpy.git}{APLpy} & Yes & \href{https://pypi.python.org/pypi/APLpy}{APLpy} & Thomas Robitaille and Eli Bressert &  \\
\href{https://github.com/astropy/astroscrappy}{Astro-SCRAPPY} & Yes & \href{https://pypi.python.org/pypi/astroscrappy}{astroscrappy} & Curtis McCully & \citealt{astroscrappy} \\
\href{http://github.com/astroML/astroML}{astroML} & Yes & \href{https://pypi.python.org/pypi/astroML}{astroML} & Jake Vanderplas &  \\
\href{https://github.com/astropy/astroplan}{astroplan} & No & \href{https://pypi.python.org/pypi/astroplan}{astroplan} & Brett Morris & \citealt{astroplan_AAS} \\
\href{http://github.com/astropy/astroquery.git}{astroquery} & Yes & \href{https://pypi.python.org/pypi/astroquery}{astroquery} & Adam Ginsburg and Brigitta Sipocz & \citealt{astroquery} \\
\href{http://github.com/astropy/ccdproc.git}{ccdproc} & Yes & \href{https://pypi.python.org/pypi/ccdproc}{ccdproc} & Steven Crawford, Matt Craig, and Michael Seifert & \citealt{ccdproc} \\
\href{https://github.com/jesford/cluster-lensing}{cluster-lensing} & No & \href{https://pypi.python.org/pypi/cluster-lensing}{cluster-lensing} & Jes Ford & \citealt{clusterlensing} \\
\href{https://github.com/adrn/gala}{gala} & Yes & \href{https://pypi.python.org/pypi/astro-gala}{astro-gala} & Adrian Price-Whelan & \citealt{gala} \\
\href{https://github.com/jobovy/galpy}{galpy} & Yes & \href{https://pypi.python.org/pypi/galpy}{galpy} & Jo Bovy & \citealt{galpy} \\
\href{http://github.com/gammapy/gammapy.git}{gammapy} & No & \href{https://pypi.python.org/pypi/gammapy}{gammapy} & Christoph Deil & \citealt{gammapy} \\
\href{http://github.com/ejeschke/ginga}{ginga} & Yes & \href{https://pypi.python.org/pypi/ginga}{ginga} & Eric Jeschke and Pey-Lian Lim & \citealt{ginga} \\
\href{https://github.com/glue-viz/glue.git}{Glue} & Yes & \href{https://pypi.python.org/pypi/glueviz}{glueviz} & Chris Beaumont and Thomas Robitaille & \citealt{glue} \\
\href{https://github.com/spacetelescope/gwcs.git}{gwcs} & No & \href{https://pypi.python.org/pypi/gwcs}{gwcs} & Nadia Dencheva & \citealt{gwcs} \\
\href{https://github.com/astropy/halotools}{Halotools} & Yes & \href{https://pypi.python.org/pypi/halotools}{halotools} & Andrew Hearin & \citealt{halotools} \\
\href{https://github.com/StingraySoftware/HENDRICS}{HENDRICS} & Yes & \href{https://pypi.python.org/pypi/hendrics}{hendrics} & Matteo Bachetti &  \\
\href{https://github.com/hipspy/hips}{hips} & No & \href{https://pypi.python.org/pypi/hips}{hips} & Christoph Deil and Thomas Boch &  \\
\href{http://github.com/spacetelescope/imexam}{imexam} & No & \href{https://pypi.python.org/pypi/imexam}{imexam} & Megan Sosey & \citealt{imexam} \\
\href{https://github.com/linetools/linetools}{linetools} & Yes & \href{https://pypi.python.org/pypi/linetools}{linetools} & J. Xavier Prochaska, Nicolas Tejos, and Neil Crighton &  \\
\href{https://github.com/Chandra-MARX/marxs}{marxs} & Yes & \href{https://pypi.python.org/pypi/marxs}{marxs} & Hans Moritz Günther & \citealt{marxs} \\
\href{https://github.com/zblz/naima}{naima} & Yes & \href{https://pypi.python.org/pypi/naima}{naima} & Victor Zabalza & \citealt{naima} \\
\href{https://github.com/RiceMunk/omnifit}{omnifit} & Yes & \href{https://pypi.python.org/pypi/omnifit}{omnifit} & Aleksi Suutarinen &  \\
\href{http://github.com/astropy/photutils.git}{photutils} & No & \href{https://pypi.python.org/pypi/photutils}{photutils} & Larry Bradley and Brigitta Sipocz & \citealt{photutils} \\
\href{https://github.com/poliastro/poliastro}{poliastro} & No & \href{https://pypi.python.org/pypi/poliastro}{poliastro} & Juan Luis Cano Rodríguez & \citealt{poliastro} \\
\href{http://github.com/weaverba137/pydl.git}{PyDL} & No & \href{https://pypi.python.org/pypi/pydl}{pydl} & Benjamin Alan Weaver & \citealt{pydl} \\
\href{https://github.com/astropy/pyregion.git}{pyregion} & Yes & \href{https://pypi.python.org/pypi/pyregion}{pyregion} & Jae-Joon Lee and Christoph Deil &  \\
\href{https://github.com/pyspeckit/pyspeckit}{pyspeckit} & Yes & \href{https://pypi.python.org/pypi/pyspeckit}{pyspeckit} & Adam Ginsburg & \citealt{pyspeckit} \\
\href{https://github.com/olebole/python-cpl}{python-cpl} & No & \href{https://pypi.python.org/pypi/python-cpl}{python-cpl} & Ole Streicher & \citealt{pythoncpl} \\
\href{https://github.com/pyvirtobs/pyvo.git}{PyVO} & No & \href{https://pypi.python.org/pypi/pyvo}{pyvo} & Stefan Becker &  \\
\href{https://github.com/astropy/regions}{regions} & No & \href{https://pypi.python.org/pypi/regions}{regions} & Christoph Deil and Johannes King &  \\
\href{https://github.com/astrofrog/reproject.git}{reproject} & Yes & \href{https://pypi.python.org/pypi/reproject}{reproject} & Thomas Robitaille &  \\
\href{http://github.com/sncosmo/sncosmo}{sncosmo} & Yes & \href{https://pypi.python.org/pypi/sncosmo}{sncosmo} & Kyle Barbary & \citealt{sncosmo} \\
\href{https://github.com/radio-astro-tools/spectral-cube}{spectral-cube} & Yes & \href{https://pypi.python.org/pypi/spectral-cube}{spectral-cube} & Adam Ginsburg & \citealt{spectralcube} \\
\href{http://github.com/astropy/specutils.git}{specutils} & No & \href{https://pypi.python.org/pypi/specutils}{specutils} & Nicholas Earl, Adam Ginsburg, Steve Crawford, Erik Tollerud &  \\
\href{https://github.com/StingraySoftware/stingray}{stingray} & No & \href{https://pypi.python.org/pypi/stingray}{stingray} & Daniela Huppenkothen, Matteo Bachetti, Abigail Stevens, Simone Migliari, and Paul Balm &  \\

      \enddata
  \end{deluxetable*}
\end{longrotatetable}

\begin{deluxetable*}{cccp{3in}c}
  \tablecaption{Registry of provisionally accepted affiliated packages.}
  \label{tab:registry_prov}
    \tablehead{
        \colhead{Package Name} &
        \colhead{Stable} &
        \colhead{PyPI Name} &
        \colhead{Maintainer} &
        \colhead{Citation}
      }
    \startdata
      \href{https://github.com/spacetelescope/spherical_geometry.git}{spherical\_geometry} & No & \href{https://pypi.python.org/pypi/spherical-geometry}{spherical-geometry} & Bernie Simon and Michael Droettboom &  \\

    \enddata
\end{deluxetable*}

\end{document}